\documentclass[aps,10pt,pra,twocolumn,amssymb,amsmath,amsfonts,showpacs,floatfix,citeautoscript,preprintnumbers,nofootinbib,superscriptaddress,a4paper]{revtex4-1}
\usepackage{graphicx,times}
\usepackage[latin1]{inputenc}

\begin{document}

\title{Hydrodynamic description of Hard-core Bosons on a Galileo ramp}

\author{Pierre Wendenbaum}
\affiliation{Institut Jean Lamour, dpt. P2M, Groupe de Physique Statistique, Universit\'e de Lorraine, CNRS, B.P. 70239, F-54506 Vandoeuvre les Nancy Cedex, France.}

\author{Mario Collura}
\affiliation{Dipartimento di Fisica dell'Universit\`a di Pisa and INFN, Pisa, Italy.}

\author{Dragi Karevski}
\email{dragi.karevski@ijl.nancy-universite.fr}
\affiliation{Institut Jean Lamour, dpt. P2M, Groupe de Physique Statistique, Universit\'e de Lorraine, CNRS, B.P. 70239, F-54506 Vandoeuvre les Nancy Cedex, France.}

\begin{abstract}
We study the quantum evolution of a cloud of hard-core bosons loaded on a one-dimensional optical lattice after its sudden release from a harmonic trap. Just after the trap has been removed,  a linear ramp potential is applied, mimicking the so called Galileo ramp experiment.  
The non-equilibrium expansion of the bosonic cloud is elucidated through a hydrodynamical description which is compared to the exact numerical evolution obtained by exact diagonalization on finite lattice sizes. The system is found to exhibit a rich behavior showing in particular Bloch oscillations of a self-trapped condensate and an ejected particle density leading to two diverging entangled condensates. Depending on the initial density of the gas different regimes of Josephson-like oscillations are observed. At low densities, the trapped part of the cloud is in a superfluid phase that oscillates in time as a whole. At higher densities, the trapped condensate is in a mixed superfluid-Mott phase that show a breathing regime for steep enough potential ramps.
\end{abstract}

\pacs{67.85.-d,03.75.Gg,03.75.Lm,67.85.Hj}

\maketitle
\section{Introduction}
It is of primary importance to understand the dynamical behavior of strongly correlated many-body quantum systems, especially since their behavior in many cases could show counter-intuitive effects with respect to the classical naive expectation. One of the early discovered such effect is the so-called Bloch oscillation regime occurring when a small constant force $F$ is applied to a quasi-particle living on a lattice \cite{Bloch1928}. In such a case, the particle momentum is drifted in time according to $q(t)=q(0)+Ft$ modulo the Brillouin zone, leading to a periodic motion with period $\tau_B=2\pi/|F|$ ($\hbar$ and the lattice spacing  are set to one). Such oscillatory behavior has been observed in many different physical contexts like semiconductors, thermal gases, photonics, cold atoms physics and with phonons \cite{BO_exp}.  At stronger forces, in multi-band systems, such oscillations can be suppressed by the Landau-Zener interband tunneling  \cite{Landau_Zener}. The Bloch oscillations can also be suppressed due to many-body effects with damping related to the integrability of the system \cite{LZ_integrable,Gustavsson}. 

On the other hand, experimental advances, notably on ultracold atomic gases have lead to a revival of the field of low-dimensional many-body systems \cite{Bloch2008}, especially within non-equilibrium aspects \cite{Dziarmaga2010,Polkovnikov2011}.  Thanks to the very low dissipation rate and very long time phase coherence of such systems, it has become possible to realize experimentally almost the  unitary dynamical evolution of ideal models like the one-dimensional Bose-Hubbard model, even in its hard-core limit \cite{Bloch2008}. 
In those systems, the release of the gas from a trap, which is a standard procedure for cold atoms experiments, can lead to interesting metastable states \cite{Heidrich-Meisner2009} or even to dynamical fermionization effects when the frequency of a parabolic trap is modified \cite{Minguzzi2005}.  Recent studies have considered the dynamical behavior of hard-core bosons in tilted optical lattices \cite{Cai2011,Collura2012},  focusing in particular on the Bloch oscillations, Landau-Zener tunneling and entanglement properties. 

In this study we pursue on the direction set by \cite{Collura2012} by considering now a hard-core boson cloud initially localized on part of an optical lattice thanks to a harmonic trap. The trap is then suddenly removed and a linear ramp potential (a constant force $F$) is added to the region initially occupied by the cloud. The unitary expansion of the cloud leads to the escape of part of the condensate while the remaining part is self-trapped into the initial region and performs Bloch oscillations. 
Thanks to the initial correlations within the cloud the escaping particles are strongly entangled with the localized oscillating condensate. 
Depending on the initial density, two different situations arise. Namely, at low density the trapped condensate is in a superfluid phase that oscillates in time as a whole while at higher density it is in a mixed superfluid-Mott phase that show a richer behavior
with, in particular, a breathing regime at large enough forces. 
The out-of-equilibrium dynamical behavior of the system is analyzed numerically by means of exact diagonalization following the methods developed in \cite{diag}  and through an analytical  hydrodynamical approximation already used in \cite{Collura2012}. 

The paper is organized as follows: in section II, we present the model, its mapping to a Fermi system and the hydrodynamical description used in this paper. In section III, we study the dynamics after the sudden release of the bosonic cloud focusing first on the emitted wave-packet, then on the self-trapped condensate and finally discussing the entanglement between the emitted particles and the self-trapped ones. In section IV, a brief summary is given.

\section{The model and its hydrodynamic description}
\subsection{The model}
We consider in this study the one-dimensional Bose-Hubbard model
describing a set of bosons leaving on a lattice with a repulsive on-site interaction $U$ and submitted to an external potential $V(t)$ which may vary in time. The dynamics is generated by the Hamiltonian  
\begin{equation*}
\mathcal{H}=-J\sum_j [a^{\dagger}_{j+1}a_j+\text{h.c.}]+\frac{U}{2}\sum_j n_j(n_j-1)+\sum_j V_j(t) n_j\; 
\end{equation*}
with $a_j$ and $a_j^\dagger$ the usual destruction and creation bosonic operators, $n_j=a_j^{\dagger}a_j$ the bosonic density at site $j$.  $J$ is the hopping magnitude and will be set to $J=1/2$ in the rest of this work (this will give a band width $\Delta=1$).  In particular,  we focus our attention on the hard-core boson limit of this model by tacking the limit $U\rightarrow \infty$. In this limit, the dynamics can be described by the new hamiltonian
\begin{equation}
\mathcal{H}=-\frac{1}{2}\sum_j [b^{\dagger}_{j+1}b_j+\text{h.c.}]+\sum_j V_j(t) n_j\; ,
\label{hamil}
\end{equation}
where the new bosonic creation and annihilation operators $b_j^{\dagger}$ and $b_j$ satisfy the on-site anticommutation rule $\{b_j,b_j^{\dagger}\}=1$,  $\{b_j,b_j\}=\{b_j^\dagger,b_j^{\dagger}\}=0$, avoiding  a double occupancy of the same site,  while  commuting at distinct sites. The standard procedure to diagonalize this Hamiltonian is to fermionise it through a Jordan-Wigner mapping and then performing a canonical Bogoliubov transformation to new diagonal Fermi operators. 
Indeed, introducing the lattice Fermi creation operators $c_j^\dagger = \prod_{i< j} (1-2n_i) b_j^\dagger$ and their adjoint annihilation operators $c_j= (c_j^\dagger)^+$, rejecting the boundaries at infinity, the Hamiltonian is expressed as a tight-binding Fermi system
\begin{equation}
\mathcal{H}=-\frac{1}{2}\sum_j [c^{\dagger}_{j+1}c_j+\text{h.c.}]+\sum_j V_j(t) n_j\; ,
\label{hamil1}
\end{equation}
with $n_j=c^{\dagger}_j c_j=b^{\dagger}_j b_j$ the occupation operator at site $j$. 
Thanks to the quadratic form of this Hamiltonian, it is readily diagonalized through a standard canonical transformation  \cite{diag} leading to 
the diagonal expression $\mathcal{H}=\sum_q \epsilon_q \eta_q^\dagger \eta_q$ where the $\epsilon_q$ are the excitation energies associated to the free Fermi particles created by $\eta_q^\dagger$ and destroyed by $\eta_q$ and the problem is in principle solved.  
However, for a quite general inhomogeneous potential $V_j(t)$ one has to compute numerically on a finite lattice the single-particle spectrum $\epsilon_q$, while  the relevant observables are expressed as determinants of the single particle Green functions 
$\langle c^\dagger_i c_j\rangle$ which have also to be computed numerically in the general case. 
Nevertheless, in the limit of a very large system with a sufficiently smooth potential function $V(x,t)$ one can well capture the features of the boson density dynamics by a hydrodynamic limit using a continuous description of the model \cite{Collura2012}. 

\subsection{Continuum limit and local equilibrium hypothesis}
\begin{figure}[ht]
\centering
\includegraphics[width=0.9\columnwidth]{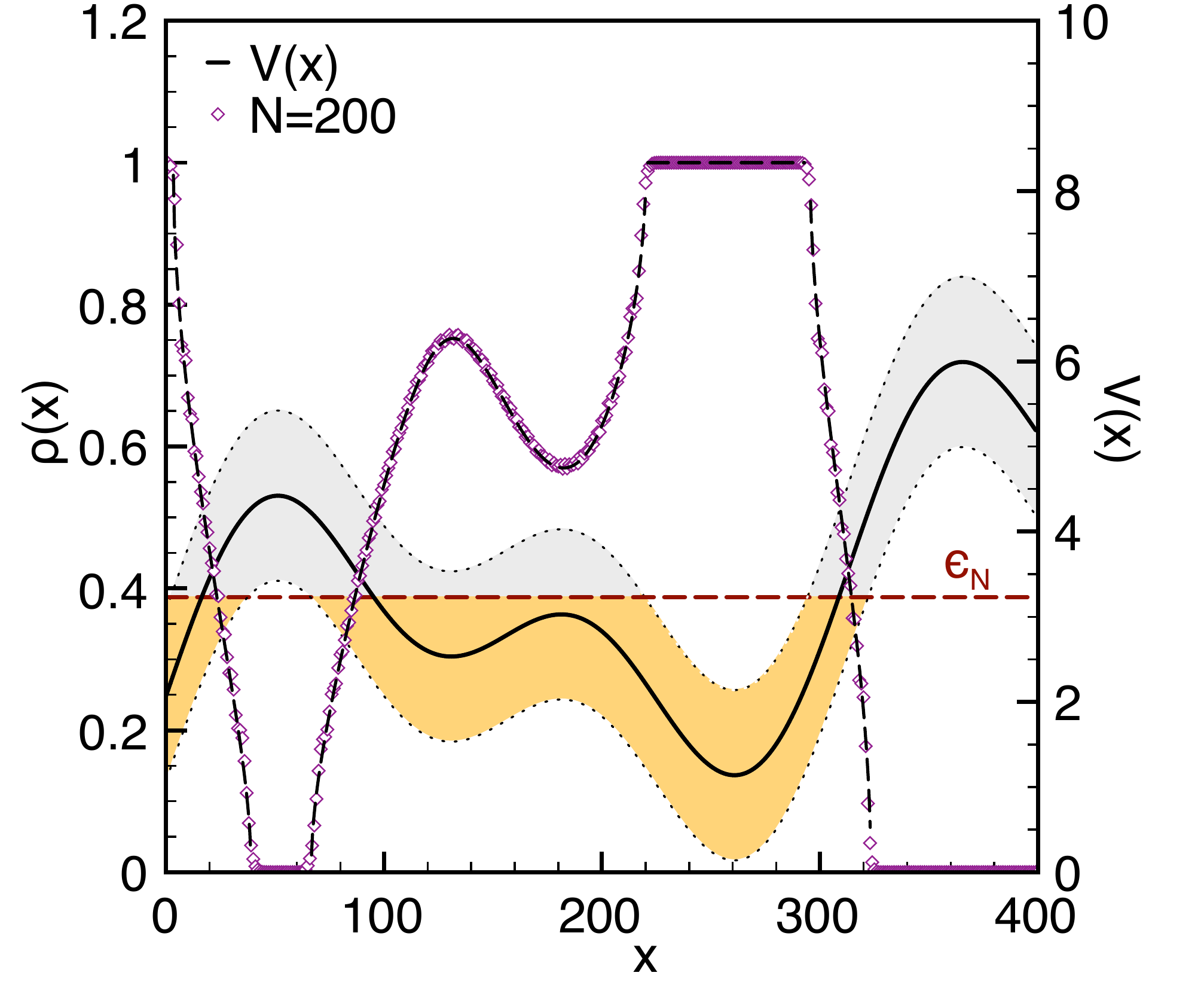}
\caption{(Color online) Ground state density profile for a system with a potential $V (x) = 2x/L+1.5 \sin(8x/L)+ \sin(16x/L)$ and $N=200$ (Exact numerical diagonalization results are represented by the symbols while the local equilibrium distribution is given by the dashed line). The local energy band is represented together with the local one-particle filled states up to the Fermi level $\epsilon_N$. 
\label{fig:local_equilibrium}}
\end{figure}

Consider the one-dimensional Hard-core bosons system on an infinite one-dimensional lattice with lattice spacing $a\ll 1$. Let the potential $V(x)$ be a smooth real function (at least let us say $V\in C^1$) on the lattice. We split the real line into regular intervals $[x, x+\Delta x]$, with $\Delta x=aM$, containing a  large number $M$ of lattice sites, while keeping the width $\Delta x$ small enough in the sense that $\forall j a\in [x,x+\Delta x],\; V(ja)\simeq V(x)$, that is the potential keeps almost a constant value on each interval.
The Hamiltonian can be recast in  the continuum limit $a\rightarrow 0$, $\Delta x\rightarrow 0$ while keeping $\Delta x/a=M\gg 1$ in the following form
\begin{equation}
\mathcal{H}=\int_{-\infty}^\infty dx\; \mathcal{H}(x)
\label{hamil2}
\end{equation}
where the hamiltonian density $\mathcal{H}(x)$ is given by 
\begin{eqnarray}
\mathcal{H}(x)&=&\frac{1}{a\Delta x}\int_0^{\Delta x} dy \left[ 
-\frac{1}{2}\Psi^\dagger (x+y) \Psi(x+y-a)+\text{h.c.} \right. \nonumber \\
 &&\left. + \Psi^\dagger(x+y) V(x) \Psi(x+y)\frac{}{} \right]
\label{hamil3}
\end{eqnarray}
in terms of the continuous creation and annihilation Fermi field operators $\Psi^\dagger(y)$ and $\Psi(y)$. 
To achieve (\ref{hamil2}) with (\ref{hamil3}) we have supposed that the interaction contribution between different intervals, which is just a local boundary term, is very small compared to the contribution  (\ref{hamil3}) within the interval. This is true if the potential variations are sufficiently small.  
The local Hamiltonian density $\mathcal{H}(x)$ is simply a continuous version of (\ref{hamil1}) with a constant potential $V(x)$, and it can be consequently easily diagonalized through the canonical mapping
\begin{eqnarray}
\Psi(x+y)=\int_0^\pi dq\; \phi_q(x+y)\eta(q,x) \; , \\
\Psi^\dagger(x+y)=\int_0^\pi dq\; \phi^*_q(x+y)\eta^\dagger(q,x) \; , 
\end{eqnarray}
where the field operators  $\eta(q,x)$ and $\eta^\dagger(q,x)$ annihilates and respectively
creates in region $[x,x+\Delta x]$ a particle with momentum $q$
 and satisfy the anti-commutation rules 
$\{\eta(p,x),\eta^\dagger(p',x')\}=\delta(p-p')\delta(x-x')$. 
The exact form of the Bogoliubov functions $\phi_q(u)$ entering into the definition of the new fields essentially depends on the boundary conditions imposed on the interval $[x, x+\Delta x]$.
Using these new fields, the total Hamiltonian takes the diagonal form
\begin{equation}
\mathcal{H}=\int_{-\infty}^\infty dx\;\int_0^\pi \; dq\;  \left[V(x)-\cos q\right] \eta^\dagger (q,x) \eta(q,x)\; .
\label{hamil4}
\end{equation}
Therefore, with respect to this Hamiltonian, the $N$-particles ground state of the bosonic system is given by a local equilibrium state. That is, all the quasi-particles associated to each phase-space points $(q,x)$, with energies $\epsilon(x, q) = V(x)-\cos q$ bellow the Fermi level $\epsilon_F(N)$, are added to the vacuum state:
\begin{equation}
|\Psi_0\rangle = \prod_{q,x}^{\epsilon_F(N)} \eta^\dagger (q,x)|0\rangle\; .
\end{equation}
The Fermi energy is given by imposing the constraint 
$\int dx\; \rho(x)=\int dx \int dq \; \eta^\dagger (q,x) \eta(q,x)=N$ on the total number of particles.
This readily implies for the ground-state density profile
\begin{eqnarray}
\rho(x)= \left\{ \begin{array}{lc}
0& V(x)-\epsilon_F>1\\
\frac{1}{\pi}\arccos (V(x)-\epsilon_F) & |V(x)-\epsilon_F|<1\\
1&V(x)-\epsilon_F<-1
\end{array} \right.
\label{density}
\end{eqnarray}
In Figure (\ref{fig:local_equilibrium}) we compare the exact numerical diagonalization with the local equilibrium prediction on a system with $L = 400$ sites at half filling and, as an illustration, a potential $V (x) = 2x/L+1.5 \sin(8x/L)+ \sin(16x/L)$. Notice the good matching between the continuum prediction (\ref{density}) and the numerical data.  In particular, we have also graphically reproduced the occupied energy levels (yellow region) at half filling factor. In practice, since $\cos q \in [-1,1]$, the energies $\epsilon(q,x)$ fall in the gray strip which follows the shape of the potential. The local equilibrium approximation breaks down whenever the local number of particle is small and the potential varies sharply.

In a grand canonical situation, when the number of particles is not fixed, the Fermi level $\epsilon_F$ in the ground state as to be set to $0$ since the ground state is build up by adding to the vacuum all excitations with negative energies :
\begin{equation}
|\Psi_0 \rangle = \prod_{\{(q,x)| \epsilon(q,x)\le 0\}} \eta^\dagger (q,x)|0\rangle\; ,
\label{initialstate}
\end{equation}
and consequently the density profile is given by (\ref{density}) with $\epsilon_F=0$ and the total number of particles in that state is just $\int dx\; \rho(x)$. 

\subsection{Initial state}
\begin{figure}[ht]
\centering
\includegraphics[width=0.45\columnwidth]{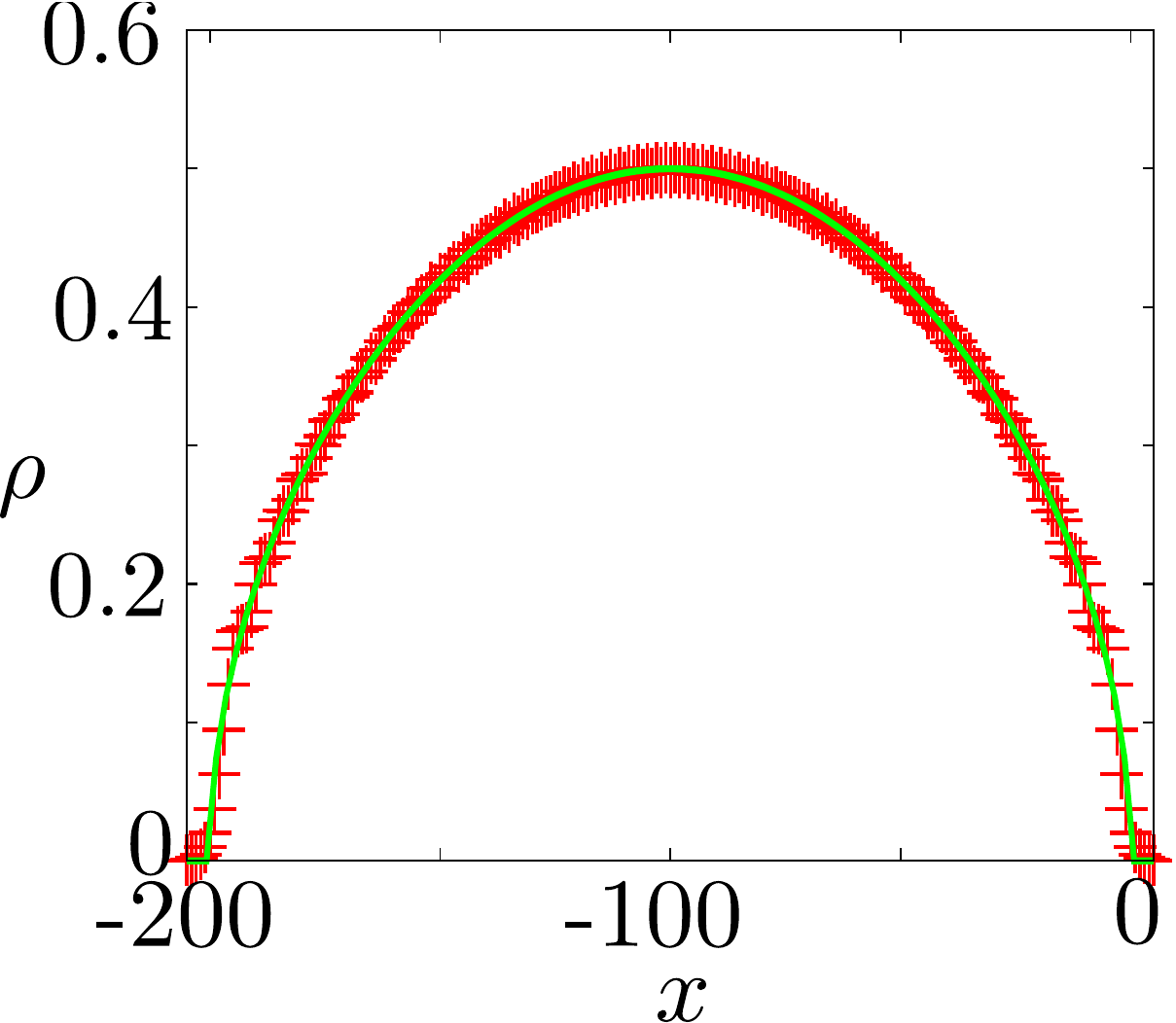}
\includegraphics[width=0.45\columnwidth]{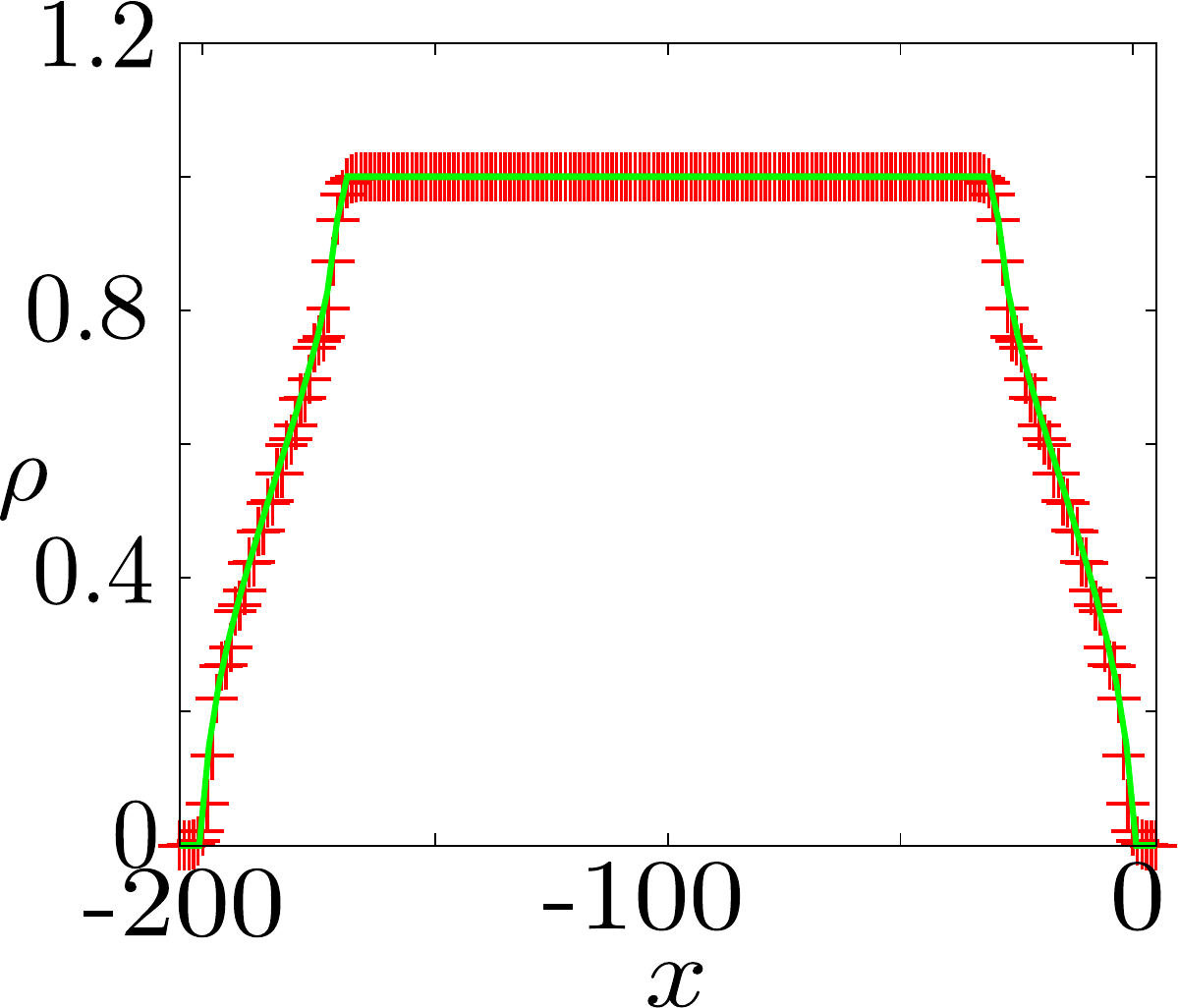}
\caption{(Color online) Left: Density profile of the initial SF state with $|A|=200$ and $\mu_0=0$. Right: Density profile of the initial SF-M state with $|A|=200$ and $\mu_0=3$.  The symbols are obtained by exact diagonalization while the full lines give the hydrodynamical profile. 
\label{fig:initialdensity}}
\end{figure}

In the following we will fix the initial state of the bosonic condensate as the grand canonical ground state associated to a harmonic trap potential.  The harmonic potential is parameterized as
\begin{equation}
V(x, t<0)=\alpha \left[x-\frac{A}{2}\right]^2-\mu_0\;, \; \alpha=\frac{4}{A^2}(1+\mu_0)\; ,
\label{potential}
\end{equation}
with $A<0$ and $\mu_0>-1$ such that there is at least few particles loaded on the lattice. 
This parametrization insures that the bosonic condensate is centered at $A/2$ with a spacial extension of width $|A|$ such that outside the region $x\in[A,0]$  the density (in the hydrodynamic limit) is vanishing (see figure~\ref{fig:initialdensity}). 
For a given width $|A|$ of the condensate, we have two qualitative distinct situations that are controlled by the value of the chemical potential $\mu_0$. Indeed, for $\mu_0<1$, the potential is never smaller than $-1$ and the condensate is fully in its superfluid phase with a density profile 
\begin{equation}
\rho(x)=\frac{\theta (-x) \theta (x+A)}{\pi} \arccos \left(\alpha \left[x-\frac{A}{2}\right]^2-\mu_0\right)\; ,
\end{equation}
where $\theta(x)$ is the Heaviside function.  We will refer to this state as the pure superfluide phase (SF-phase). 
On the other hand, for $\mu_0>1$, there is in the middle of the condensate a Mott insulating phase, with $\rho=1$, that extends spacially over the region $x\in [\frac{A}{2}-\Delta_{Mott}, \frac{A}{2}+\Delta_{Mott}]$ with 
\begin{equation}
\Delta_{Mott}=\frac{|A|}{2} \sqrt{\frac{\mu_0-1}{\mu_0+1}}\; .
\end{equation}
This Mott phase is surrounded by two superfluid phases, of spatial extensions 
\begin{equation}
\Delta_{SF}=\frac{|A|}{2}- \Delta_{Mott}=\frac{|A|}{2}\left( 1-\sqrt{\frac{\mu_0-1}{\mu_0+1}} \right)\; ,
\end{equation}
with a profile given by (\ref{density}) with the harmonic potential (\ref{potential}), as illustrated on figure~\ref{fig:initialdensity}. As the chemical potential $\mu_0$ is getting larger and larger, the Mott phase is growing at the expense of the superfluid phases, shrinking them to the boundaries close to $x=A$ and $x=0$. We call this state the mixed superfluid-Mott phase (Mixed SF-Mott phase).

\section{Dynamics after the sudden quench}

\subsection{Sudden release of the trap and loading of the linear ramp}
At time $t=0^+$, starting from the previous initial state, we suddenly release the gas from the parabolic trap and load an external constant force $F$ on the negative side of the real axis. The force is described by the linear ramp potential 
$$
V(x,t>0)=-F x \; \theta(-x) 
$$
and the system dynamics is governed by a new Hamiltonian (\ref{hamil1}) with potential $V(x,t>0)=-F x \; \theta(-x)$. 
As we will see in the following, the main features of the dynamics are well described in the hydrodynamic limit. 
As seen previously, the initial state has approximately a coarse-grained phase-space density 
$$
w(x,q)= \frac{1}{\pi}\theta(-x)\theta(x+A) \theta(q)\theta(q_F(x)-q)
$$
with a local Fermi wave-vector  $q_F(x)=\pi \rho(x)$ where $\rho(x)$ is given by (\ref{density}). The local initial density is then simply given by adding all the local quasiparticles (associated to each phase-space point) with different wave-vectors $q$: $\rho(x)=\int dq\; w(x,q)$. 
Just after the sudden quench, the energy of a particle initially at position $x$ is shifted by the additional potential energy $V (x,t>0)$. This brutal change of potential leads, as a consequence, to a unitary out-of-equilibrium evolution of the condensate 
from its initial state. In order to understand that evolution, taking into account that the dynamics is  unitary, one has to realise that the initial quasiparticles that build up the local density are emitted to the right and to the left on trajectories of constant energies. This approach will allow us to reconstruct the evolving density profile $\rho(x; t)$ and the associated current density profile $j(x;t)$.

\subsection{Wave-packet emission}
\subsubsection{Density profile}
\begin{figure}[t]
\centering
\includegraphics[width=1\columnwidth]{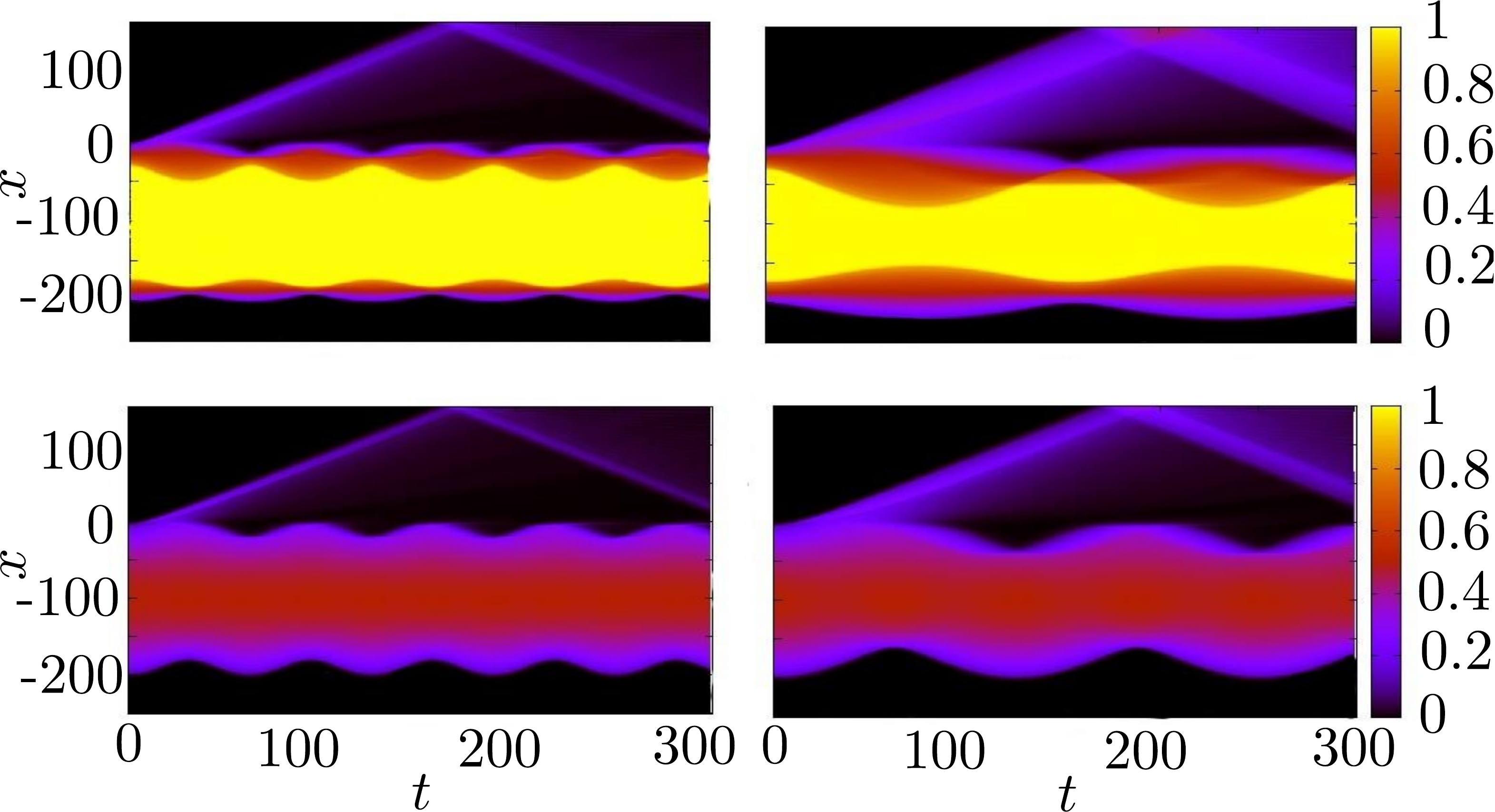}
\caption{(Color online) Snapshots of the evolution of the density profile obtained from exact numerical calculation for the two different initial states (SF-Mott and SF phases) at two different forces $F$. Up-left: SF-Mott phase with $F=0.1$. Up-right: SF-Mott phase with $F=0.04$. Down-left: SF phase with $F=0.1$. Down-rigth: SF phase with $F=0.05$. 
\label{fig:density}}
\end{figure}

\begin{figure}[t]
\centering
\includegraphics[width=0.4\columnwidth]{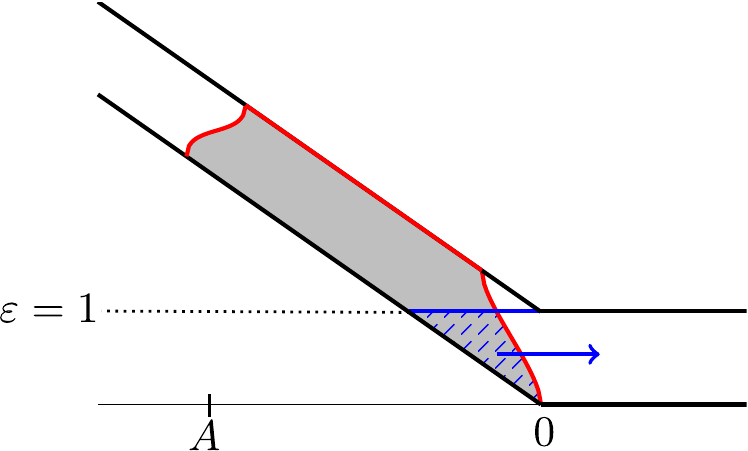}
\includegraphics[width=0.55\columnwidth]{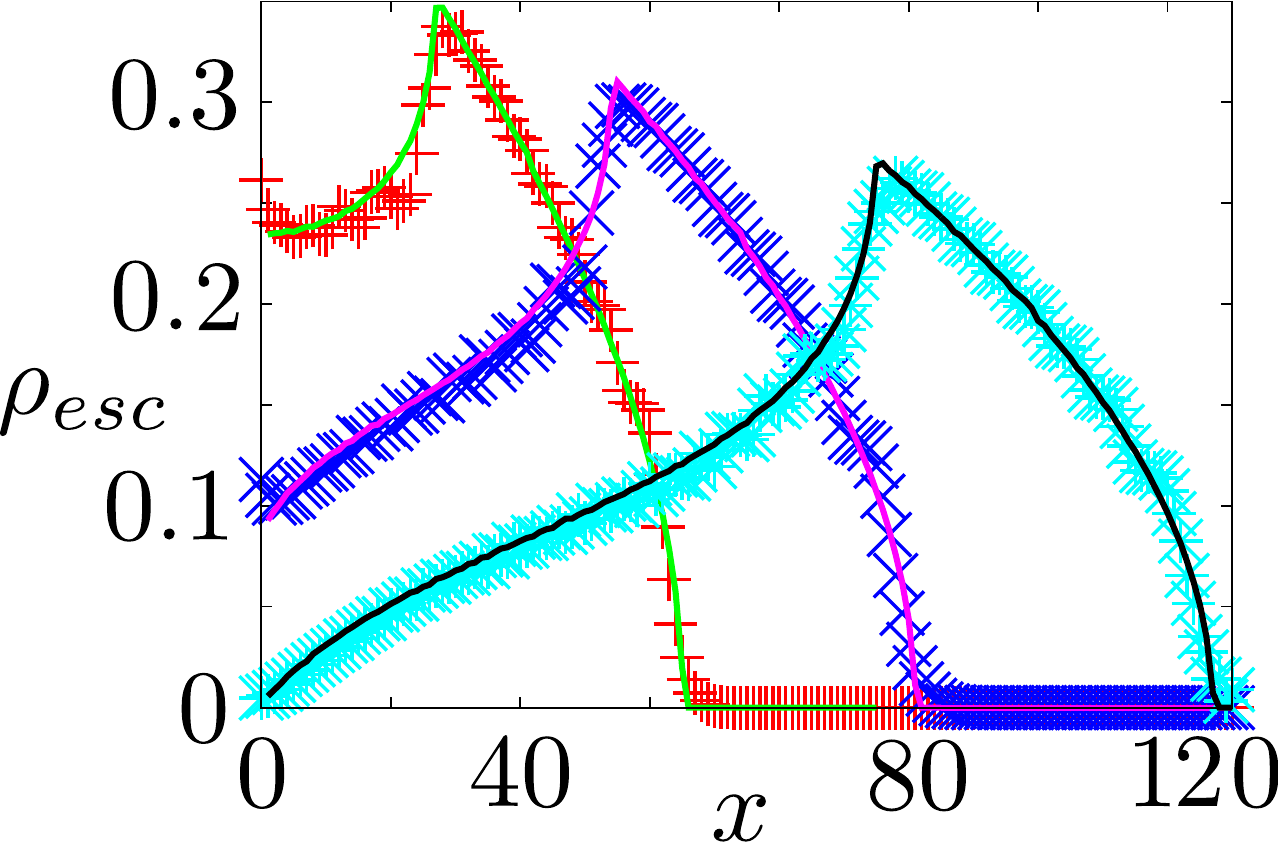}
\caption{(Color online) \textbf{(a)} Schematic picture of an  initial SF-M state just after the sudden quench of the linear ramp. The dashed blue region represents the density that will be ejected into the propagative band. 
\textbf{(b)}  Density profile of the ejected particles at times, from left to right, $t=74$, $t=110$ and $t=156$ . The symbols represent the exact diagonalization results obtained on a chain of 400 sites with a force $F=0.04$ and with an initial SF-M state compared to the hydrodynamic predictions (full lines).
\label{fig:escape}}
\end{figure}

The typical evolution of the bosonic density, just after the sudden quench, is shown on figure~\ref{fig:density} for several forces and for the two distinct typical initial states, that is the pure SF-phase and the mixed SF-Mott state. 
As we see clearly on figure~\ref{fig:density} only part of the bosons are ejected to the right with a corresponding density that is spreading in time (notice that since the numerical results are obtained on a finite lattice with open boundary conditions the ejected particles are reflected when they reach the right boundary wall).  The remaining particles are self-trapped into the initial region, performing Bloch oscillations (we will come back on this point in the next section). 
The reason for the escape of the particles to the right is best understood from the hydrodynamical description. Indeed, the population living initially on the tilted band with energies within the interval $[-1,1]$ are connected at $t=0^+$ to the propagating states of the energy band at $x>0$ (see figure~\ref{fig:escape}.a)). Those particles will consequently escape from there initial position and propagate to the right toward $+\infty$. Right movers will escape directly while left movers will  be first reflected on the left tilted band edge and then propagate toward the right direction. From energy conservation, within the tilted band, as for Bloch oscillations, the momentum of a right(left) mover with energy $\epsilon$ evolves as $q^\pm(t)=\pm q +Ft$, describing the conversion of the potential energy $V(x)$ into kinetic energy  $-\cos q(t)$. As the escaping left and right movers have reached the origin $x=0$, all the potential energy has been converted into kinetic energy. This happens at times $t^\pm=\frac{1}{F}\left(\mp q + \arccos(-\epsilon)\right)$ for the right($+$ sign) and the left($-$ sign) movers respectively. In the region $x>0$, the particles evolve at constant velocity $v(\epsilon)=\sqrt{1-\epsilon^2}$ depending only on there (initial) energy $\epsilon$. Consequently the initial right and left movers are traveling along the trajectories 
\begin{equation}
x^\pm (\epsilon, q, t)= \sqrt{1-\epsilon^2}\left( t-t^\pm\right)
\end{equation}
which can be written in terms of the initial position $x$ as 
\begin{eqnarray}
&x^\pm (x, q, t)&=\sqrt{1-(Fx+\cos q)^2}\nonumber \\
&\times& \left( t-\frac{1}{F}\left( \mp q +\arccos(Fx+\cos q)\right) \right).
\end{eqnarray}
The density profile of the escaping particles is then given by the sum of the left and right movers contributions
\begin{equation}
\rho_{\text{esc}}(x;t)=\rho_{\text{esc}}^+(x; t)+\rho_{\text{esc}}^-(x; t)
\end{equation}
with
\begin{eqnarray}
\rho_{\text{esc}}^\pm(x; t)=\int_{-1}^1 \frac{d\epsilon}{F}\int_{Q(\epsilon)} \frac{dq}{2\pi} \delta(x-x^\pm (\epsilon, q, t))
\label{rhoesc}
\end{eqnarray}
where the domain of integration $Q(\epsilon)$  over the momenta is shown on figure~\ref{fig:escape}.a). 
In figure~\ref{fig:escape}.b) we show the exact numerical results obtained by exact diagonalization of the escaping density compared to the hydrodynamic prediction (\ref{rhoesc}). We see a very good agreement up to small interference effects that are obviously neglected in the continuum limit. It is interesting to notice that the rightmost front of the escaping wave-pacquet shows, beyond the hydrodynamic envelop, a staircase structure as already noticed in a different context~\cite{Hunyadi2004,Platini2005}. Each of the stairs corresponds to a particle, the integrated density over each such stair being equal to one, ejected to the right and moving ballistically with velocity $v(\epsilon)=\sqrt{1-\epsilon^2}$. Behind the front the exact structure is a bit more complicated and it is difficult to discriminate between different particles.    

\subsection{Dynamics of the trapped particles}
\subsubsection{Density profile}
As seen from the snapshot \ref{fig:density}, due to energy conservation, there is a self-trapped density in the initial region which exhibits a quite rich behavior with Bloch oscillations related to the fact that the particles are traveling on a lattice and experience a constant force $F$ \cite{Bloch1928}. This leads to a momentum drift in time according to $q(t)=q(0)+Ft$ ($\hbar$ and the lattice spacing are set to one) modulo the Brillouin zone and then leading to a periodic motion with period $\tau_B= 2\pi/F$. It is also remarkable to notice that while in the case of a pure SF initial state the condensate oscillates in time as a whole (both sides of the condensate are moving in phase) while in the case of a mixed SF-Mott initial state one has to distinguish between a low force regime and a high force one.  Indeed, at low forces (see $F=0.04$ in figure \ref{fig:density}) the two superfluid phases surrounding the Mott phase are oscillating in phase (both moving on the same direction at a given time) while at high forces  ($F=0.1$ in figure \ref{fig:density}) the two superfluid parts are moving in opposite directions at a given time, leading to a breathing condensate. 
This behavior is easily understood thanks to the hydrodynamical approach where 
the density profile of the trapped particles is given by the sum of left and right movers densities:
\begin{equation} 
\rho(x;t)=\rho^+(x; t)+\rho^-(x; t)\; .
\label{trapped-density}
\end{equation}
The left and right movers densities are given by
\begin{equation}
\rho^\pm(x,t)=\int_0^\pi \frac{dq}{2\pi} \Pi_{[f_2(q),f_1(q)]}(g^\pm(x,q,t))
\end{equation}
where
\begin{equation}
g^\pm(x,q,t)=x-\frac{1}{F}[\cos q- \cos(q\pm Ft)]
\end{equation}
and
\begin{eqnarray}
f_1(q)&=&\left\{\begin{array}{lr}
-\frac{1+\cos q}{F}&\quad q\in[0,\tilde{q}]\\
\frac{A}{2}+\sqrt{\frac{\cos q +\mu_0}{\alpha}} &\quad q\in [\tilde{q},\pi]
\end{array}\right.\\
f_2(q)&=& \frac{A}{2}-\sqrt{\frac{\cos q +\mu_0}{\alpha}}
\end{eqnarray}
with $\tilde{q}=\arccos (-1-Fx_M)$ where $x_M$ is the right-most locus of the initial trapped condensate (see figure \ref{fig:escape}.a)).

\begin{figure}[t]
\centering
\includegraphics[width=0.95\columnwidth]{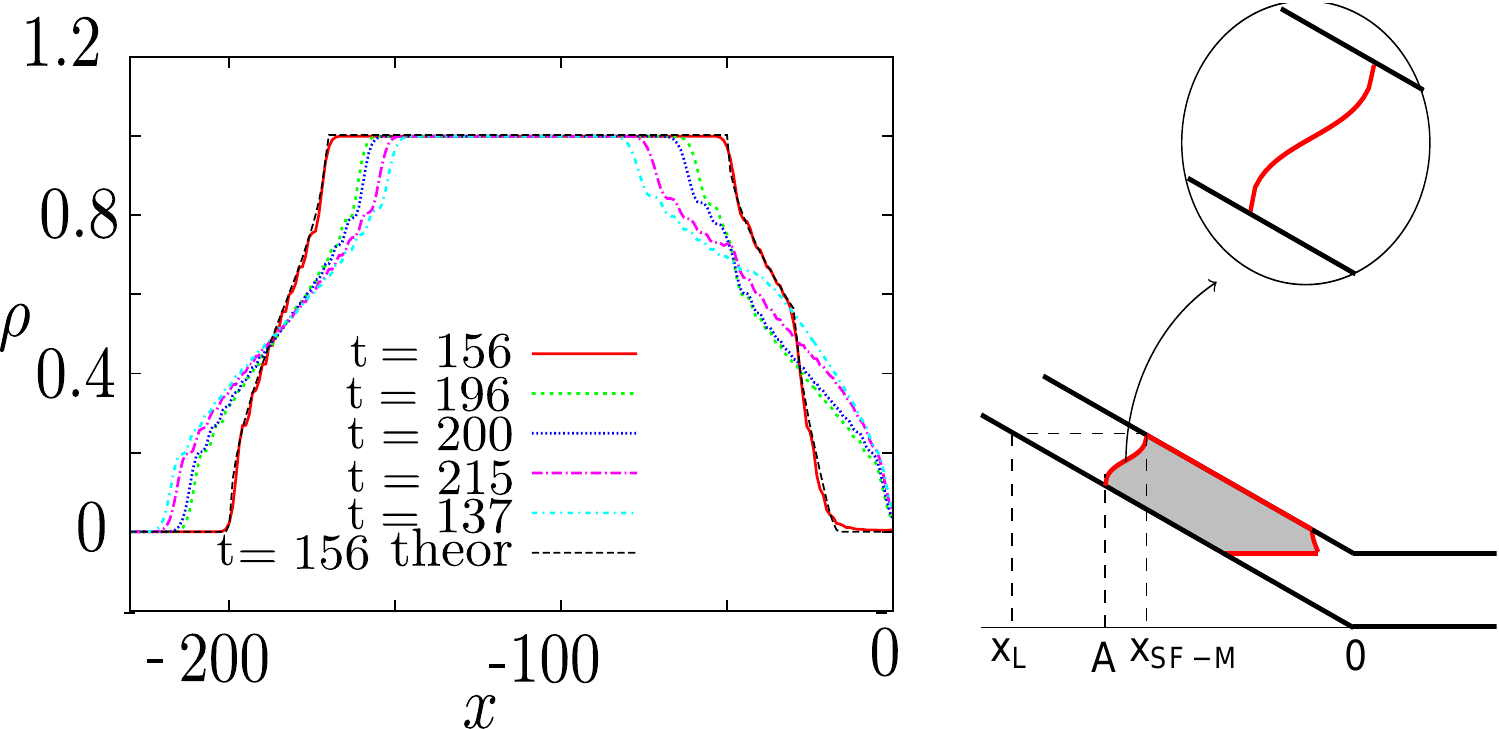}
\includegraphics[width=0.95\columnwidth]{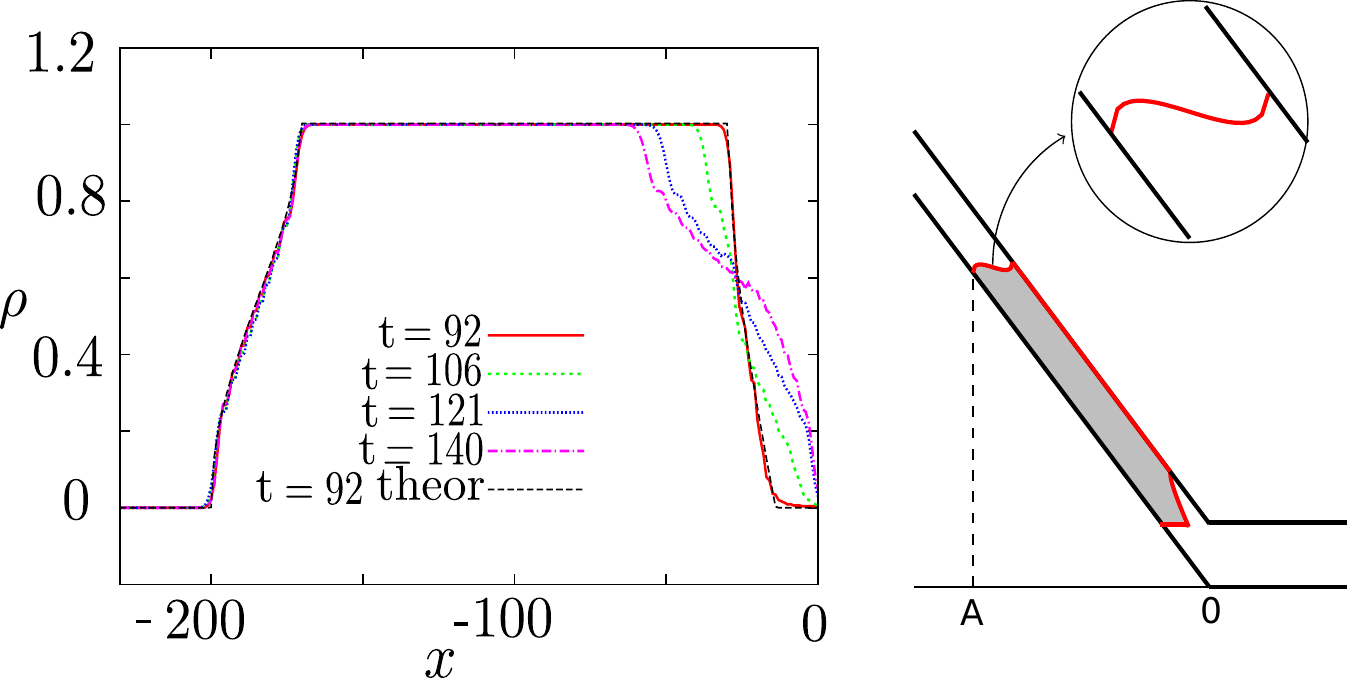}
\includegraphics[width=0.95\columnwidth]{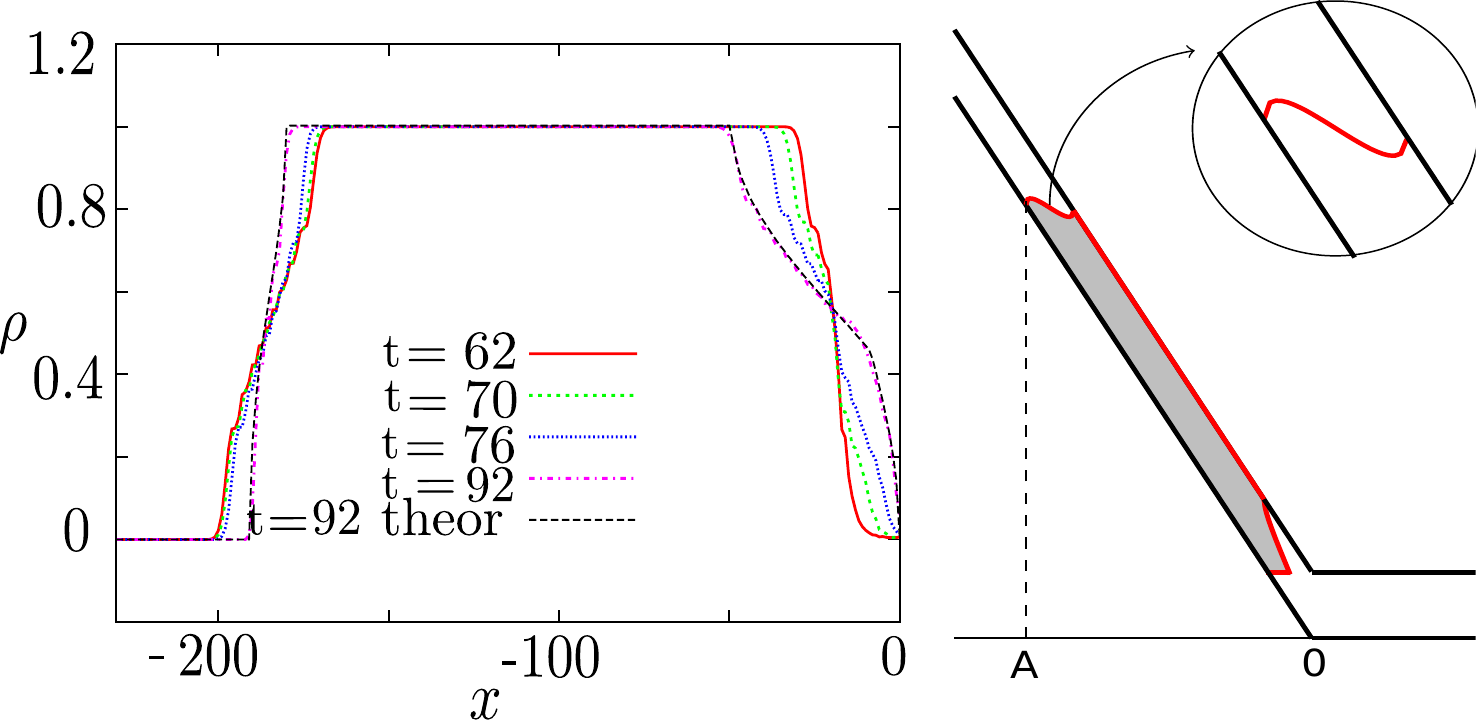}
\includegraphics[width=0.95\columnwidth]{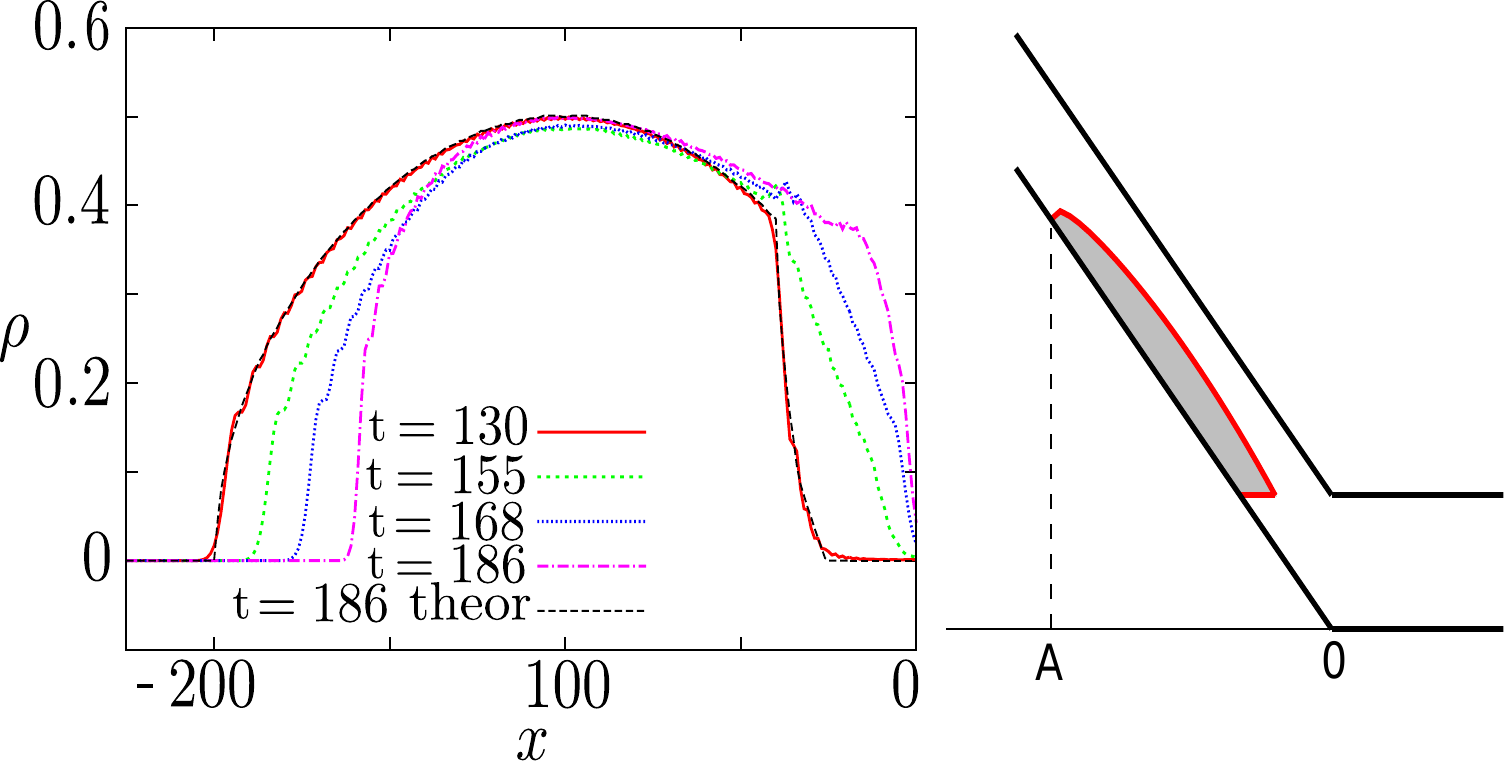}
\caption{(Color online) \textbf{(a)} Top: Self-trapped density profile for an initial SF-Mott state at different times for $F=0.04$. The hydrodynamical prediction is given by the dashed line. 
\textbf{(b)} Middle-up: Same as \textbf{(a)} for $F^*=0.0689.$. 
\textbf{(c)} Middle-down: Same as \textbf{(a)} for $F=0.1$. 
\textbf{(d)} Bottom: Self-trapped density profile for an initial SF state at different times for $F=0.05$. The hydrodynamical predictions are represented in each case by the dashed lines. 
\label{fig:densities}}
\end{figure}

On figure \ref{fig:densities} we have plotted over a half-period the oscillating density profile of the self-trapped particles for the two initial states (SF-Mott and SF phases) as extracted from the exact numerical diagonalization results shown in figure \ref{fig:density} and compared to the hydrodynamical prediction (\ref{trapped-density}), together with a schematic representation of the initial state. First of all we see a very good agreement of the hydrodynamical prediction with the exact numerical results. 
In these figures we clearly see the two different oscillation regimes in the SF-Mott case (see figures \ref{fig:densities}.a) and  \ref{fig:densities}.c)). The reason for that is easily understood from the schematic pictures of the initial states. 
Indeed at low enough forces, as seen on the schematic picture of the initial state, the energy of the local density at the separation point $x_{SF-M}$ between the SF phase and the Mott phase is higher  than the energy at the left-most initial locus $x=A$ and consequently the density will explore regions on the left of $A$ up to the locus $x_L=x_{SF-M} -2/F$. This phenomenon disappears when the energy at the locus $x_{SF-M}$ is getting smaller than the energy at the point $x=A$.  This precisely appears above a threshold force $F^*$ given by $-F^*A-1= -F^*x_{SF-M} +1$.    
In the case represented in figure \ref{fig:densities}.b) the threshold force is given by $F^*=0.0689.$ where we see almost no evolution of the profile at the left side. Beyond that value, as seen on figure \ref{fig:densities}.c) for $F=0.1$, the left and right sides of the density profile are moving in phase.   
With a SF initial state, the self trapped density is always globally oscillating as a whole (see figure \ref{fig:densities}.d)). 

\subsubsection{Current density profile}
\begin{figure}[hbt]
\centering
\includegraphics[width=0.7\columnwidth]{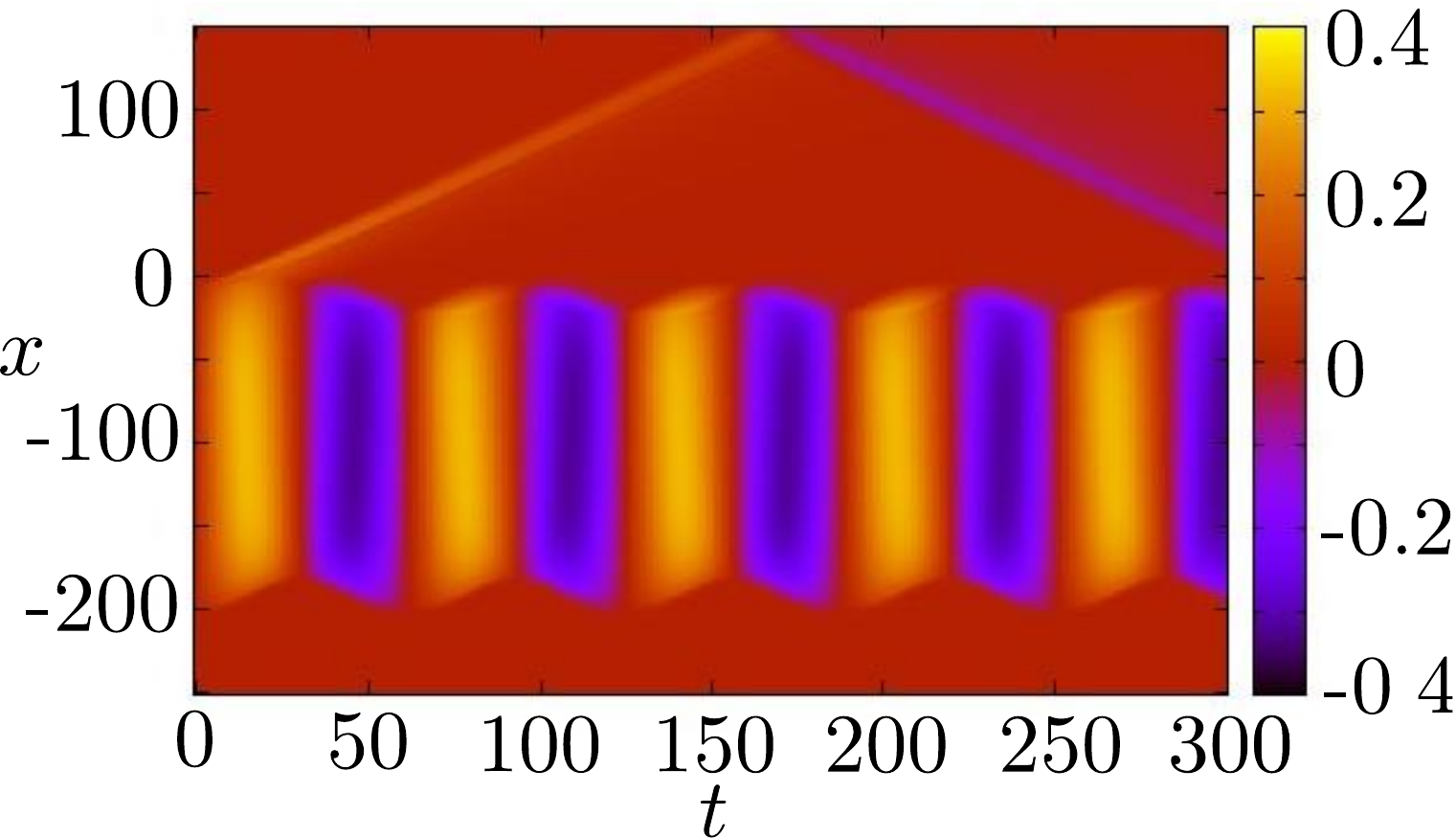}
\includegraphics[width=0.7\columnwidth]{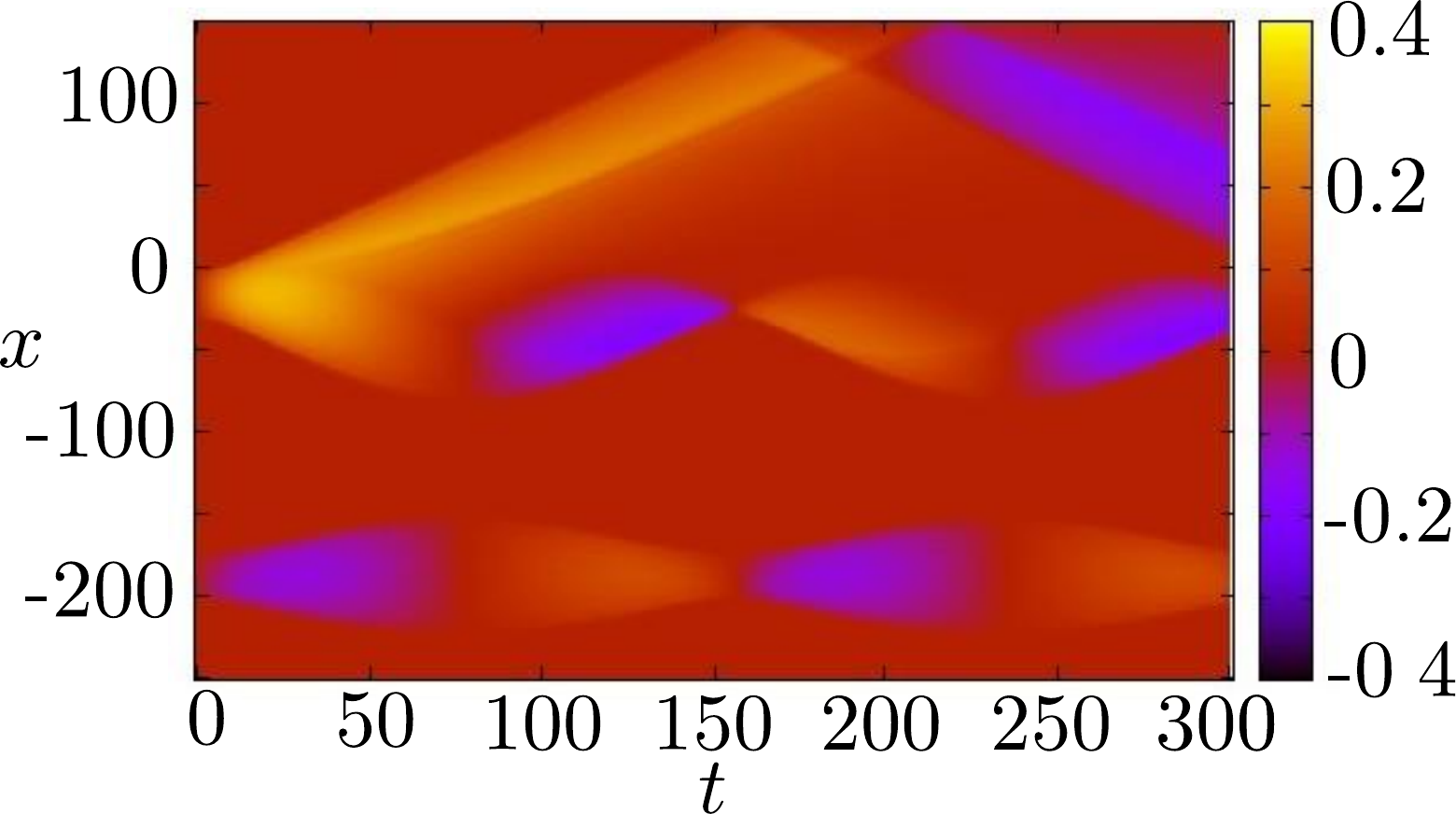}
\includegraphics[width=0.7\columnwidth]{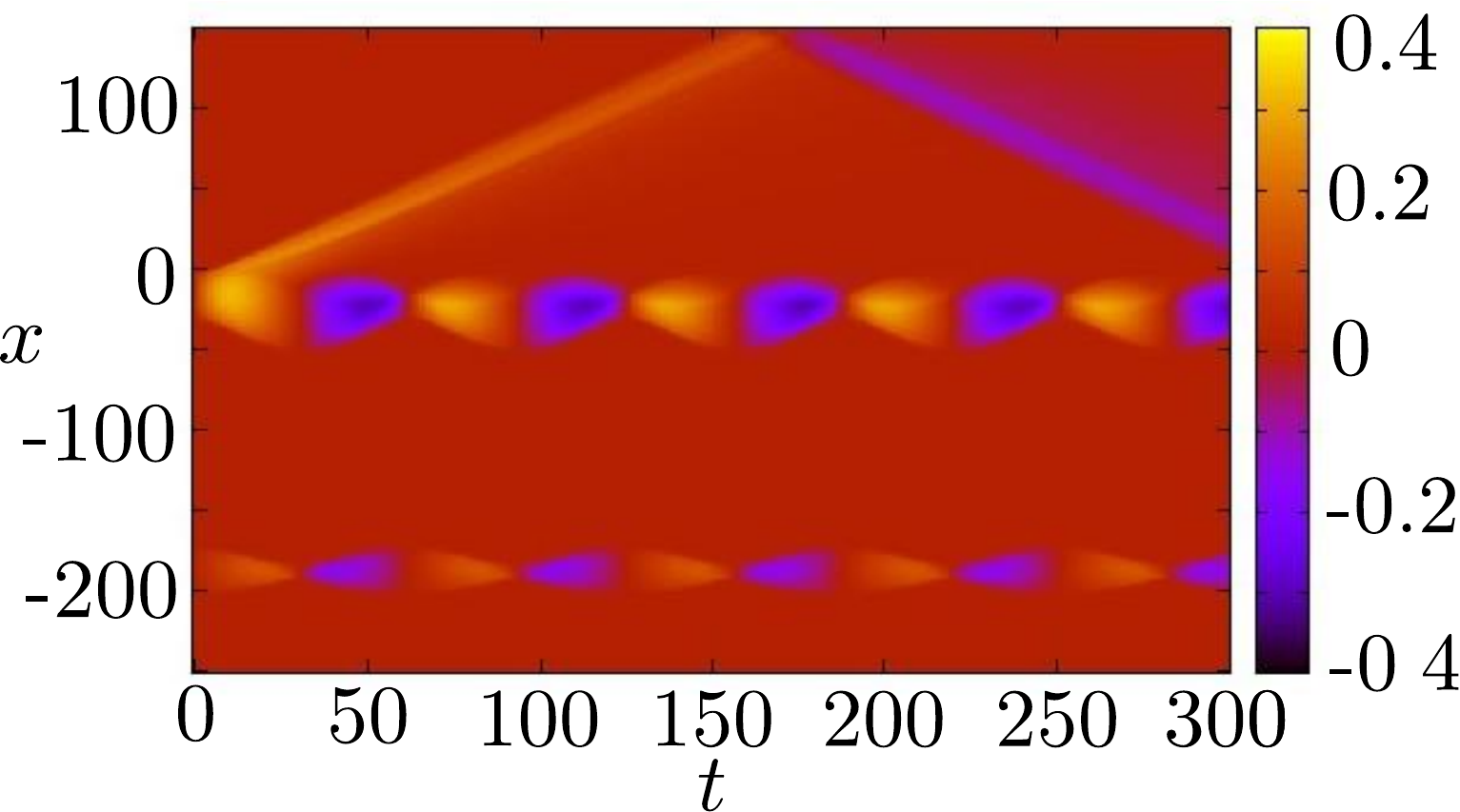}
\caption{(Color online) Snapshot of the time evolution of the current density profile for a SF initial state at $F=0.05$ (top) and a SF-Mott initial state at low forces $F=0.04$ (middle) and high forces $F=0.1$ (bottom). 
\label{fig:current}}
\end{figure}

This remarkable periodic motion of the trapped density is naturally associated to a flow of particles giving rise to a periodic current density $j(x,t)=\langle J(x,t) \rangle$ which is defined through the continuity equation $i[H,n(x)]=-\nabla J(x)$ where $n(x)$ is the occupation operator at site $x$. 
In figure \ref{fig:current} we show a snapshot of the current density obtained from exact diagonalization  for a SF state at $F=0.05$ (top) and a SF-Mott initial state at low forces ($F=0.04$) and high forces ($F=0.1$).  
In both cases, in the trapped region we see clearly the Bloch oscillations with a strip structure for the pure SF phase indicating the collective motion of the superfluid condensate while in the mixed SF-Mott situation we observe the oscillations (in phase at $F=0.04$ and in opposite phase at $F=0.1$) of the two SF phases surrounding a stationary Mott plateau.

\begin{figure}[hbt]
\centering
\includegraphics[width=1\columnwidth]{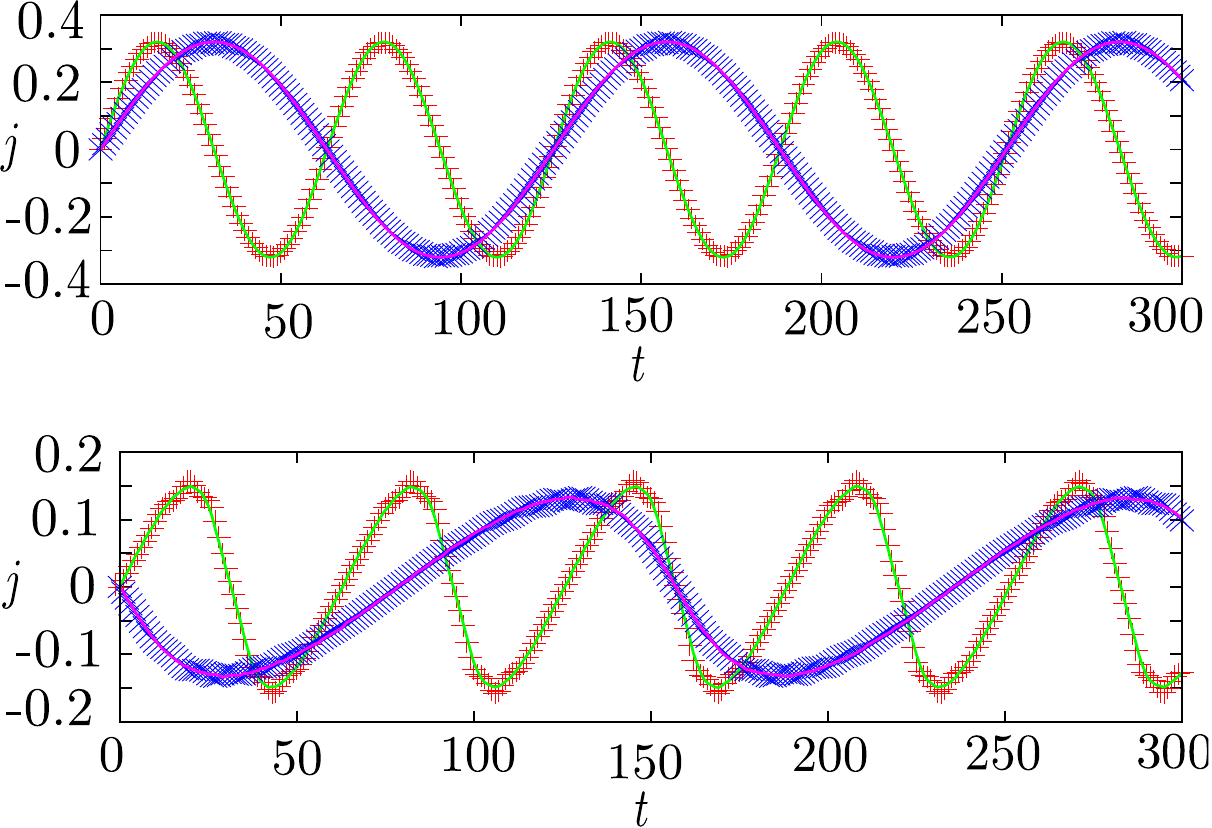}
\caption{(Color online) Top: Current density at $x=-100$ for $F=0.1$ (shortest period) and $F=0.05$ in the SF case. 
Bottom: Current density at $x=-185$ for $F=0.1>F^*$ (shortest period) and $F=0.04<F^*$ in the SF-Mott case. Full lines are the corresponding hydrodynamical predictions. 
\label{fig:currentheo}}
\end{figure}

In the hydrodynamical limit, the current density of the self-trapped condensate is simply given by summing over all the quasiparticles current contributions $\rho(q) v(q)$ with the velocities $v(q)=\pm \sin (q(0)\pm Ft)$ leading to 
\begin{eqnarray}
&j(x,t)&=\int_0^\pi \frac{dq}{2\pi} \sin(q+Ft) \Pi_{[f_2(q),f_1(q)]}(g^+(x,q,t)) \nonumber \\
&-&\int_0^\pi \frac{dq}{2\pi} \sin(q-Ft) \Pi_{[f_2(q),f_1(q)]}(g^-(x,q,t))\; .
\label{current}
\end{eqnarray}
We have plotted in figure \ref{fig:currentheo} the current density (\ref{current}) compared to the exact numerical one at different positions as a function of time for the two cases, SF and SF-M. We observe a perfect matching of the hydrodynamical prediction (\ref{current}) with the exact numerical values. We also observe a very nice sinus behavior of the current in the middle of the superfluid condensate reminiscent of a Josephson-type oscillations which is explicitly given from (\ref{current}) as $j(A/2,t)=1/\pi \sin(Ft)$.

\subsection{Entanglement between the trapped and the escaping particles}
Just after the sudden unloading of the parabolic trap and the quench of the linear ramp, the initial condensate is split into two disjoint parts: the escaping particles and the remaining self-trapped ones. Due to the initial correlations in the starting state (\ref{initialstate}), these two well separated condensates are entangled. This entanglement between propagative modes (into the right propagative energy band) and bound states (into the self-trapping region) can be  quantified through the bi-partite von Neuman entropy 
$S(x,t)=-\text{Tr}\{\rho(x,t)\ln \rho(x,t)\}$
where $\rho(x,t)=\text{Tr}_{y>x} \{|\Psi(t)\rangle \langle \Psi(t)|\}$ is the reduced density matrix associated to the left of position $x$ on the lattice, deduced from the time-evolved pure state $|\Psi(t)\rangle$ by tracing out the degrees of freedom on the right of $x$ (see for example the review \cite{Calabrese}).

\begin{figure}[bt]
\centering
\includegraphics[width=1\columnwidth]{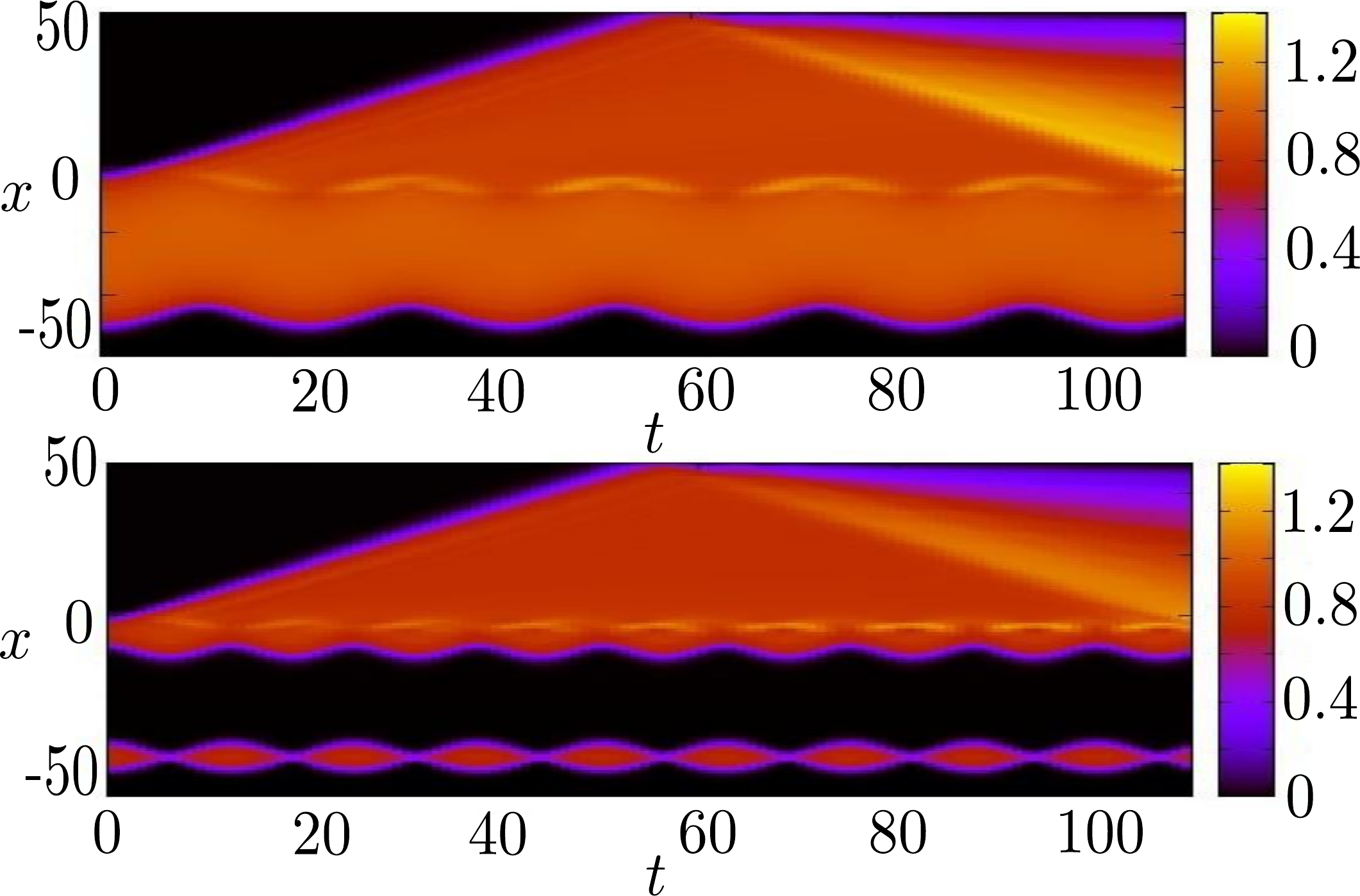}
\caption{(Color online) Top: Entanglement entropy as a function of space and time for $F=0.3$ with a SF initial state generated with $A=50$ and $\mu_0=0$.
Bottom: Entanglement entropy as a function of space and time for $F=0.5$ with a SF-M initial state generated with $A=50$ and $\mu_0=2$.
\label{fig:entropy}}
\end{figure}

\begin{figure}[hbt]
\centering
\includegraphics[width=0.48\columnwidth]{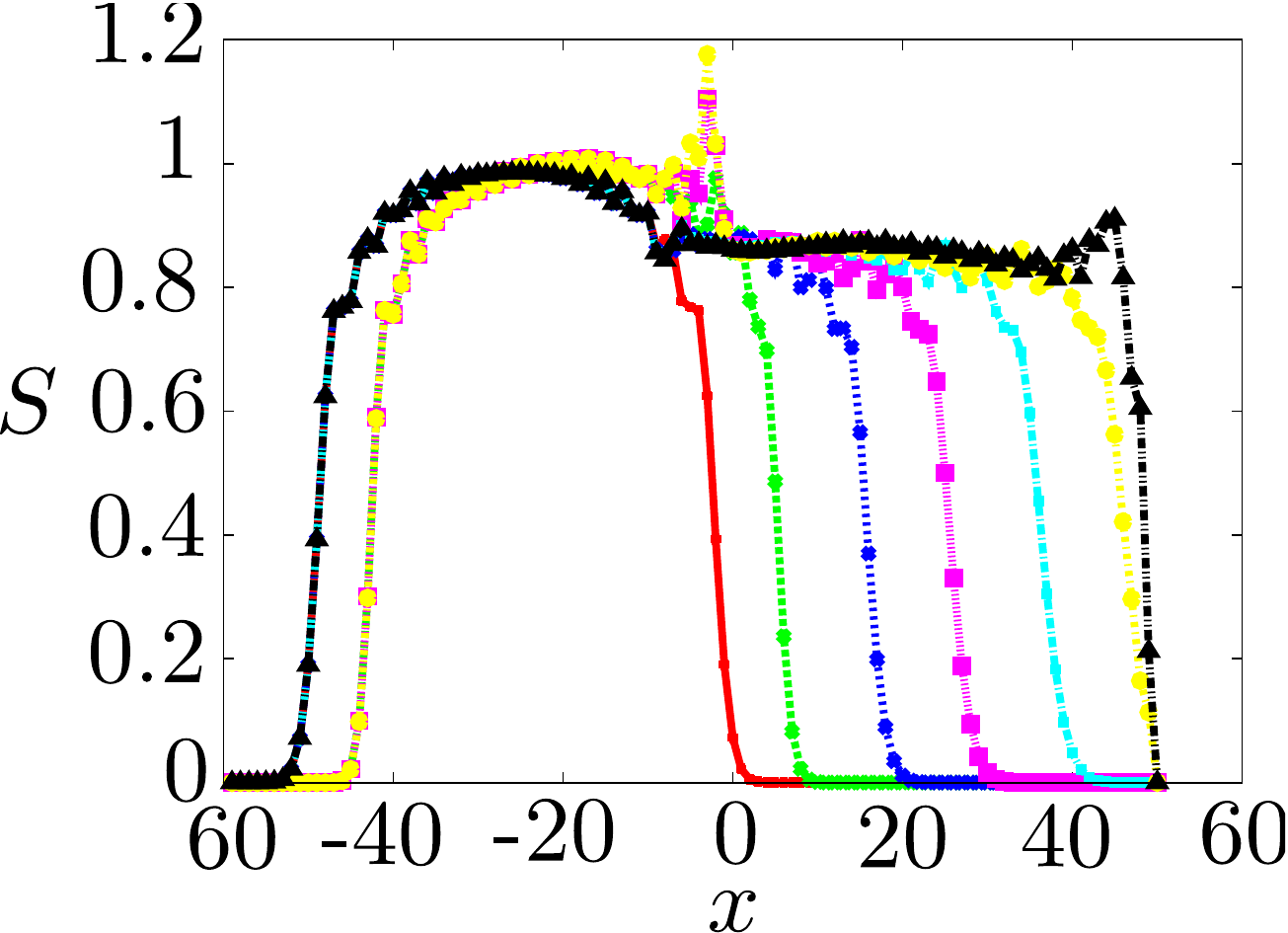}
\includegraphics[width=0.48\columnwidth]{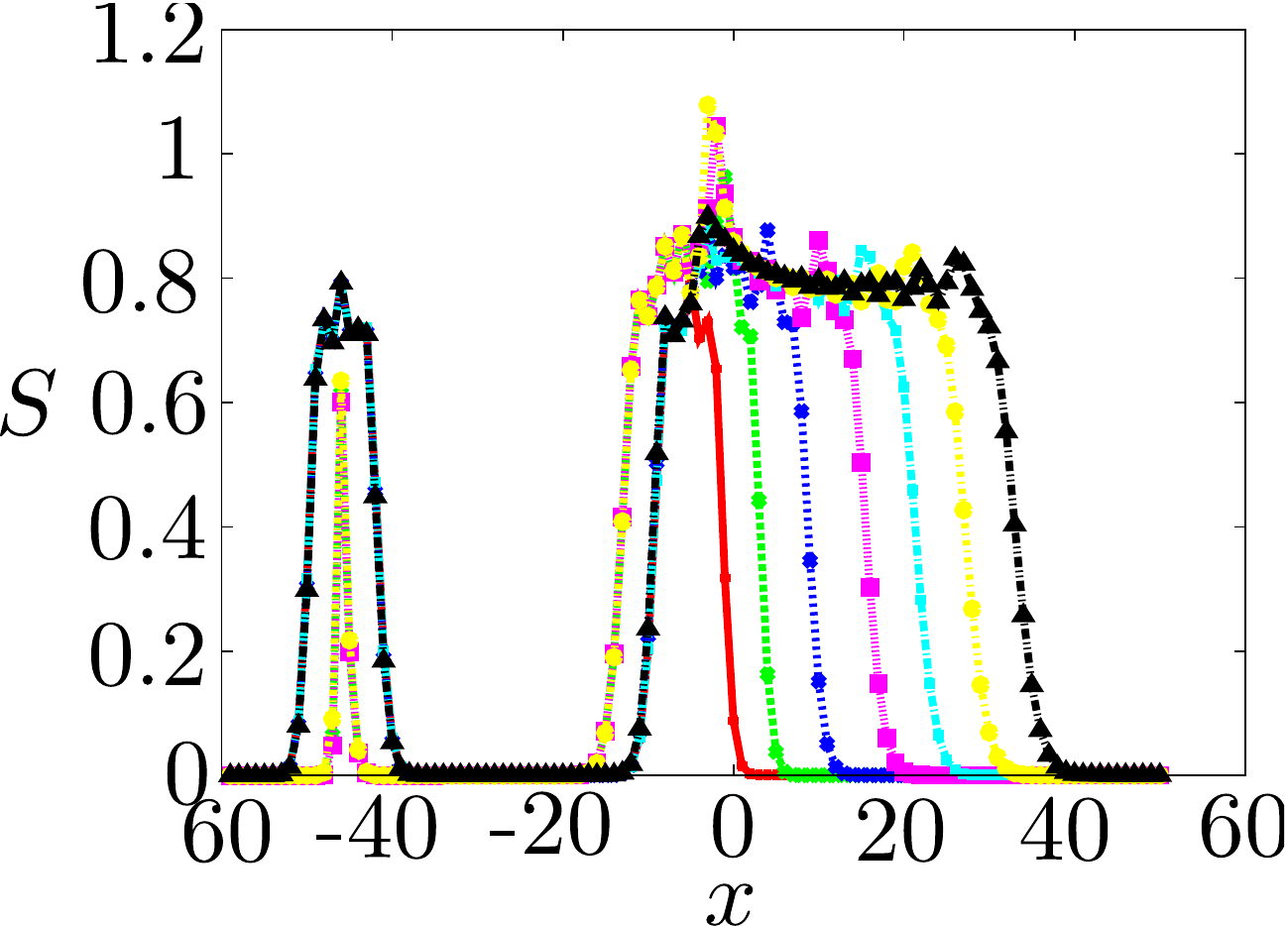}
\caption{(Color online) Left: Entanglement entropy profile at different multiples of the Bloch period $\tau_B=2\pi/F$ and half-period  $\tau_B/2$  for $F=0.3$ with a SF initial state generated with $A=50$ and $\mu_0=0$.
Right: Entanglement entropy profile at different multiples of the Bloch period $\tau_B=2\pi/F$ and half-period  $\tau_B/2$  for $F=0.5$ with a SF-M initial state generated with $A=50$ and $\mu_0=2$.
\label{fig:entropy2}}
\end{figure}

As seen on figure \ref{fig:entropy} we have qualitatively two different situations depending on whether the initial state is a pure SF state (see the top snapshot of  figure \ref{fig:entropy})  or a mixed SF-Mott one (see the other snapshot of figure \ref{fig:entropy}). 
First of all, in both cases we see clearly the entanglement generated between the self-trapping region and the propagative band by the ballistic motion of the quasi-free particles leaving the initial region. We also clearly see that in the SF-M case, the entanglement is vanishing in the Mott phase since there the state is a direct product of single occupancy local states. 
Notice also that the initial entanglement within the self-trapped condensate is basically conserved in time, showing a trivial time-evolution related to the periodic oscillations of the trapped condensate. Indeed, in figure \ref{fig:entropy2} we have plotted the entanglement entropy profile at integer multiples of the Bloch period $\tau_B=2\pi/F$ and of the half-period to show the very nice superposition of the profiles for the trapped particles. Basically, it means that the initial correlations within the trapped wave-packet remain unchanged during the time evolution and no entanglement is lost or created.

\section{Conclusion}
We studied the dynamical behavior of a hard core boson gas initially prepared in a parabolic trap and subject to the sudden quench of a linear ramp potential just after the release of the trapping potential (like in the so-called Galileo ramp experiment with classical particles). The study was based on exact numerical diagonalization methods and on a hydrodynamical description which allowed the very good understanding of the dynamical  behavior at the level of local density and local current density.  The dynamics of this setup is rich, showing an escape of particles   and Bloch oscillations for the self-trapped remaining condensate. Depending on the initial trap, the state of the condensate is either in a purely  superfluid phase or in a mixed state with a Mott phase surrounded by two superfluid phases.  In these two cases the behavior of the self-trapped condensate is different, showing in particular in the SF-M phase the possibility for a breathing condensate at high forces while in the SF situation the condensate is always oscillating as a whole.   Notice finally that this setup can be used to create entangled many-body wave-packets by coupling the Bloch bound states to  propagative modes.

This work is supported by ANR-09-BLAN-0098-01. M.~C. and P.~W. benefited from the support of
the International Graduate College on Statistical Physics and Complex
Systems between the universities of Lorraine and Leipzig.

\end{document}